# Unlocking AI Data Center Interconnection Capacity Through Coordinated Grid and Data Center Flexibility

Rida Fatima and Xingpeng Li

*University of Houston, 4222 Martin Luther King Blvd, Houston, 77204, Texas, USA*

## Abstract

The rapid growth of AI data centers is creating concentrated electricity demands that can exceed available distribution network headroom and delay interconnection. This paper investigates whether capacity can be used effectively by coordinating flexibility on both sides of the interconnection. A day-ahead mixed integer second order cone programming framework maximizes feasible AI data center IT capacity using utility side conservation voltage reduction and network topology reconfiguration, together with data center workload shifting and reserve constrained uninterruptible power supply storage. Background feeder demand uses voltage dependent ZIP models, while the data center is modeled as constant power demand using Training, Inference, and Mixed workload profiles. A two-stage solution first maximizes interconnection capacity and then minimizes feeder losses for the retained capacity, with numerical relaxation screening. Studies on a 116-bus model derived from the IEEE 123-node feeder show that, at the constrained bus-60 connection point, grid side flexibility, driven almost entirely by network reconfiguration, increases feasible IT capacity by approximately 42.5-45.0%, while data center flexibility alone provides approximately 2.1-5.4% gains. Coordinated operation increases feasible IT capacity from approximately 10.5-10.6 MW under BASE operation to 15.3-16.0 MW, corresponding to gains of approximately 45.9-50.8%; the 16.0 MW Inference result is limited by the adopted UPS reserve requirement. Benefits are strongly location dependent and influenced by workload flexibility, deferral duration, PUE, UPS energy and reserve requirements, and feeder loading. Under a combined adverse operating condition, FULL capacity decreases by only 3.4% relative to the nominal Mixed case while retaining substantial additional capacity over BASE.

## Nomenclature

Sets:

| | |
|---|---|
| $S$ | Set of Substations |
| $N$ | Set of Nodes |
| $E$ | Set of Branches |
| $E^{sw}$ | Set of switchable branches |
| $T$ | Set of scheduling intervals |

Indices:

| | |
|---|---|
| $s$ | Substation $s$, an element of $S$ |
| $n$ | Node $n$, an element of set $N$ |
| $e = (i, j)$ | Branch $e$ connecting nodes $i$ and $j$, $e \in E$ |
| $t$ | Time $t$, an element of set $T$ |

Parameters:

| | |
|---|---|
| $V_s^{base}$ | Base feeder head voltage |
| $V_s^{CVR,min}$ | Minimum allowable feeder head voltage under CVR |
| $P_s^{sub,max}$ | Maximum allowable active power supplied by a substation |
| $Q_s^{sub,max}$ | Maximum allowable reactive power supplied by a substation |
| $r_{ij}$ | Resistance of branch $(i, j)$ |
| $x_{ij}$ | Reactance of branch $(i, j)$ |
| $V_n^{rated}$ | Rated voltage magnitude at node $n$ |
| $V_n^{min}$ | Minimum allowable nodal voltage magnitude |
| $V_n^{max}$ | Maximum allowable nodal voltage magnitude |
| $I_{ij}^{max}$ | Maximum allowable current magnitude of branch |
| $M_{ij}$ | Big M parameter for the switch dependent voltage drop constraint |
| $A^{max}$ | Maximum number of switching actions over the scheduling horizon |
| $K_Z^P$ | Active load constant impedance |
| $K_I^P$ | Active load constant current |
| $K_P^P$ | Active load constant power |
| $K_Z^Q$ | Reactive load constant impedance |
| $K_I^Q$ | Reactive load constant current |
| $K_P^Q$ | Reactive load constant power |
| $L_t^{DC}$ | Normalized data center workload profile at time $t$ |
| $PUE$ | Power usage effectiveness of the data center |
| $\varphi$ | Fraction of IT workload that is flexible |
| $J_{ij}^0$ | Initial status of controllable branch $(i, j)$ |
| $S_{ij}^{max}$ | Apparent power rating of branch$(i, j)$ |
| $H$ | Maximum allowable workload deferral duration |
| $P_{max}^{UPS,flex}$ | Maximum UPS charging or discharging power available for grid support operation |
| $E_{max}^{UPS}$ | Energy capacity of the dedicated UPS |
| $\eta^{ch}$ | UPS charging efficiency |
| $\eta^{dch}$ | UPS discharging efficiency |
| $SOC^{init}$ | Initial state of charge of the UPS |
| $h^{res}$ | Required UPS emergency reserve duration |
| $\rho^{crit}$ | Fraction of peak facility demand requiring emergency backup |
| $N^{cyc}$ | Maximum allowable daily UPS discharge throughput in equivalent cycles |
| $\Delta t$ | Duration of each interval $t$ |
| $P_n^0$ | Nominal active background demand at node $n$ |
| $Q_n^0$ | Nominal reactive background demand at node $n$ |
| $L_t^{BG}$ | Hourly background load multiplier |
| $\lambda^{BG}$ | Background load scaling factor |

Variables:

| | |
|---|---|
| $v_{n,t}$ | Squared voltage magnitude at node $n$ and time $t$ |
| $C^{DC}$ | Installed IT power capacity referenced to the peak of the unshifted normalized workload profile. |
| $\ell_{ij,t}$ | Squared current magnitude on branch $(i,j)$at time $t$ |
| $J_{ij,t}$ | Binary status of branch $e$ at time $t$, equal to 1 if closed and 0 if open. |
| $a_{ij,t}$ | Switching action indicator for branch $e$ at time $t$ |
| $P_{ij,t}$ | Power flow on branch $e$ at time $t$ |
| $Q_{ij,t}$ | Reactive power flow on branch $e$ at time $t$ |
| $X_{\tau,t}$ | Flexible IT workload arriving at time $\tau$ and executed at time $t$ |
| $P_t^{IT}$ | Actual scheduled IT power demand at time $t$ |
| $P_t^{FAC}$ | Total data center facility power demand at time $t$ |
| $P_{n,t}^{DC}$ | Net active power drawn by the data center from the distribution network |
| $P_t^{ch}$ | UPS charging power at time $t$ |
| $P_t^{dch}$ | UPS discharging power at time $t$ |
| $E_t^{UPS}$ | Stored energy of the dedicated UPS at time $t$ |
| $u_t^{ch}$ | Binary UPS charging status at time $t$ |
| $u_t^{dch}$ | Binary UPS discharging status at time $t$ |
| Auxiliary Expressions: | |
| $P_{n,t}^{BG}$ | Voltage dependent active background demand |
| $Q_{n,t}^{BG}$ | Voltage dependent reactive background demand |
| $\tilde{V}_{n,t}$ | Approximated voltage magnitude term for the constant current ZIP component |
| $P_{n,t}^{sub}$ | Active power supplied by substation at node $n$ |
| $P_{n,t}^{PV}$ | Net PV active power injected at node $n$ |
| $P_{n,t}^{gen}$ | Net dispatchable generation at node $n$ |
| $P_{n,t}^{BESS}$ | Net feeder BESS active power at node $n$ |
| $\Delta C^{int}$ | Interaction between grid side and data center side flexibility |
| $\bar{P}_t^{IT}$ | Baseline IT demand before workload scheduling |

## 1. Introduction

Growth in AI computing is accelerating the deployment of large data centers whose concentrated electrical demand is creating new challenges for power system planning and interconnection. Training and deploying modern AI models requires substantial computing infrastructure, and continued growth in model size, inference activity, and digital services is accelerating the development of new data center capacity. The International Energy Agency projects substantial growth in global data center electricity consumption through 2030, while recent United States assessments similarly indicate significant increases in both annual electricity use and peak demand [1]-[4]. This growth is increasingly concentrated in regions where multiple large facilities compete for limited transmission and distribution capacity. Consequently, access to available grid capacity, utility interconnection timelines, and required network upgrades are becoming important constraints on data center development [5], [6].

The electrical demand of a data center consists of the information technology (IT) equipment together with cooling, power conversion, lighting, and other supporting infrastructure. Power Usage Effectiveness (PUE), defined as the ratio of total facility power to IT equipment power, is widely used to characterize data center energy efficiency [7]. A lower PUE therefore allows a larger portion of the available electrical capacity to be used directly by computing equipment. In addition to facility efficiency, data center electrical demand depends on the underlying computational workload. AI training typically involves sustained computational activity, whereas inference demand may exhibit greater temporal variation because it is influenced by user requests and service activity. Mixed facilities can contain characteristics of both workload types.

Publicly available high resolution electrical traces that separately characterize different AI computing workloads remain limited. Synthetic load generation tools therefore provide a useful alternative for system planning studies when their assumptions and processing steps are clearly reported. In this work, representative data center load profiles are constructed using the Lawrence Berkeley National Laboratory Data Center and Industrial Electrical Load Shape Maker [8]. Annual synthetic load series are used to identify representative 24-hour weekday profiles for Training, Inference, and Mixed workload conditions. The selected profiles are normalized to their respective daily peaks, placing the workload archetypes on a common installed IT capacity basis.

Unlike many conventional electrical loads, part of the computing demand of a data center can also exhibit temporal flexibility. Delay tolerant computing jobs may be shifted within acceptable execution windows without curtailing the total computational work. Previous studies have investigated data center flexibility through temporal workload scheduling, spatial workload migration, demand response, and coordinated energy management [9]-[14]. Cao et al. [10] quantified data center flexibility associated with periodic batch workloads, while Chen et al. [11] demonstrated that controllable data center demand can improve the operational flexibility of active distribution networks. Other studies have developed coordinated scheduling frameworks that link computing workloads with energy systems and electricity demand [12]-[14]. These investigations establish that data centers can operate as partially flexible electricity consumers rather than completely rigid loads.

Data centers also contain energy resources that may contribute to short-term electrical flexibility. Uninterruptible power supply systems and their associated batteries are primarily installed to maintain critical computing services during supply interruptions. However, part of this stored energy can potentially support grid operation when sufficient emergency reserve is retained. Recent studies have examined the use of data center batteries for energy flexibility and demand response while recognizing the reliability constraints associated with their primary backup function [15], [16]. Broader studies have also identified workload scheduling, battery storage, backup generation, cooling systems, and geographical workload migration as potential sources of data center flexibility [15]-[19]. These findings motivate the consideration of both computing flexibility and battery support when evaluating the electrical interconnection of future AI data centers.

### *1.1 Data Center Hosting Capacity and Distribution Network Flexibility*

While much of the existing literature considers the operation of data centers after they have been connected, the rapid growth of AI infrastructure has increased attention toward the amount of new data center demand that existing networks can accommodate. Fang and Li [20] investigated hosting capacity for edge data centers in active distribution networks and demonstrated that feasible capacity can vary significantly with connection location and network constraints. That work focuses on hosting capacity assessment and enhancement for edge data centers under active distribution network constraints. In contrast, the present study explicitly coordinates feeder side CVR and NTR with temporal workload scheduling and reserve constrained UPS operation to quantify the interconnection headroom created by flexibility on both sides of a prescribed connection point. At the transmission level, Manassero et al. [21] considered data center allocation while accounting for transmission hosting capability and resilience. Samir et al. [22] examined the hosting capacity of AI training data centers under frequency related constraints. Collectively, these studies show that data center capacity is not determined solely by the facility demand itself but also by the available electrical headroom, network operating conditions, and point of interconnection.

Distribution systems already contain operational controls that can modify this available headroom before conventional network reinforcement is undertaken. Conservation voltage reduction (CVR) reduces the consumption of voltage sensitive loads by operating feeder voltages toward the lower portion of their acceptable range. Numerous studies have demonstrated the potential of CVR for demand and energy reduction in distribution systems [23]-[25]. However, the effectiveness of CVR depends strongly on the voltage sensitivity of the underlying demand, making appropriate load representation important when evaluating its benefits [23]-[26]. ZIP load models are particularly useful because they distinguish constant impedance, constant current, and constant power components of feeder demand.

Network topology reconfiguration (NTR) provides a different form of operational flexibility. NTR changes the status of sectionalizing and tie switches to redistribute feeder power flows while maintaining an acceptable radial network structure. Reconfiguration has traditionally been applied for loss reduction, voltage improvement, congestion relief, and load balancing [27]-[29]. More recent studies have coordinated network reconfiguration with voltage control and other distribution resources, demonstrating that topology and voltage decisions can interact through their effects on feeder demand and power flows [30].

These grid side controls are particularly relevant for data center interconnection because they affect network headroom through different physical mechanisms. CVR can reduce the voltage sensitive portion of the surrounding feeder demand, thereby releasing part of the existing capacity. NTR can redistribute power flows and source utilization across different feeder sections and substations. The data center itself, however, is dominated by power electronic equipment and should not be assumed to exhibit the same voltage dependent behavior as conventional feeder loads. Treating the data center separately as constant power demand therefore allows the contribution of CVR to be attributed to reductions in background demand rather than to an artificial reduction in computing requirements.

Existing studies nevertheless tend to examine grid flexibility and data center flexibility separately. Data center energy management research generally focuses on workload scheduling, demand response, storage operation, electricity cost, or ancillary services after a facility has been interconnected [9]-[19]. Hosting capacity research primarily evaluates connection location, network constraints, reinforcement requirements, or additional electrical resources [20]-[22]. At the same time, distribution flexibility research investigates controls such as CVR and NTR primarily for conventional load management, loss reduction, or voltage regulation [23]-[30]. The potential value of coordinating these resources, specifically for AI data center interconnection, has received comparatively limited attention.

### *1.2 Research Gap and Contributions*

An important gap therefore exists between data center flexibility studies and distribution network hosting capacity studies. Limited attention has been given to the interconnection problem in which operational flexibility available on both sides of the connection point is coordinated to determine how much additional AI computing capacity can be accommodated by existing distribution infrastructure.

On the utility side, CVR can reduce voltage sensitive background demand while multi substation NTR can redistribute feeder flows and available source capacity. On the data center side, workload scheduling can reduce the coincidence between computing demand and network constraints, while a dedicated UPS battery can temporarily modify grid demand subject to emergency reserve requirements. These flexibility resources operate through fundamentally different mechanisms, and their combined value may depend on data center workload characteristics, connection location, network loading, facility efficiency, and backup requirements. The extent to which grid side and data center flexibility provide distinct, overlapping, or additional interconnection capacity benefits has not been sufficiently quantified in absolute capacity terms.

To address this gap, this paper develops a day ahead mixed integer second order cone programming framework that maximizes the feasible peak IT capacity of an AI data center while coordinating utility side and data center side flexibility. The framework considers CVR and multi substation NTR on the distribution side, together with workload scheduling and reserve constrained UPS battery operation on the data center side. Background feeder demand is represented by using voltage dependent ZIP models, while the AI data center is modeled separately as constant power demand. A two-stage solution procedure first maximizes feasible IT interconnection capacity and subsequently minimizes feeder active energy losses while retaining the accepted capacity and discrete operating decisions.

The main contributions of this work are summarized as follows:

1. A coordinated AI data center interconnection framework is developed to maximize feasible peak IT capacity at a prescribed connection location by jointly using utility side CVR and multi substation NTR with data center workload scheduling and reserve constrained UPS operation.
2. Conventional feeder demand and AI data center demand are modeled separately using voltage dependent ZIP loads and constant power demand, respectively. This prevents CVR from artificially reducing computing demand and enables its actual contribution to interconnection headroom to be quantified.
3. Separate mechanism cases and a descriptive difference in differences capacity metric are used to summarize the individual and combined effects of grid side and data center side flexibility.
4. The robustness and practical limits of the resulting interconnection capacity are evaluated through sensitivities to connection location, workload flexibility, deferral duration, PUE, UPS sizing, emergency reserve, and background feeder loading, together with an adverse operating condition test, numerical screening, and independent AC power flow validation.

The proposed framework therefore treats operational flexibility as a source of interconnection headroom rather than only as a mechanism for reducing operating cost after a data center has already been connected. It provides a systematic approach for identifying when operational flexibility can increase the amount of AI computing demand accommodated at a prescribed, potentially constrained point of interconnection before conventional network reinforcement is required.

## 2. Proposed Modeling and Optimization Framework

This study develops a day-ahead MISOCP framework to determine the maximum AI data center capacity that can be accommodated by an active distribution network over a 24 h horizon. Grid side flexibility is provided by conservation voltage reduction (CVR) and network topology reconfiguration (NTR), while data center flexibility is provided by workload scheduling and a dedicated UPS battery. Existing PV, dispatchable generation, and feeder BESS remain available within their operating limits. Background feeder demand is represented using voltage dependent ZIP loads, whereas the data center is modeled separately as constant power demand so that CVR does not directly reduce computing demand.

### *2.1 Distribution Network Model*

The distribution network is represented by using the branch flow formulation for radial systems [31]-[34]. For branch $e = (i, j)$ and time $t$, squared voltage and current variables are defined as

$$v_{i,t} = |V_{i,t}|^2$$
$$\ell_{ij,t} = |I_{ij,t}|^2$$

The active and reactive branch flows are denoted by $P_{ij,t}$ and $Q_{ij,t}$, respectively. For a closed branch, the voltage relation is

$$v_{i,t} - v_{j,t} = 2\left(r_{ij}P_{ij,t} + x_{ij}Q_{ij,t}\right) - \left(r_{ij}^2 + x_{ij}^2\right)\ell_{ij,t} \tag{1}$$

where $r_{ij}$ and $x_{ij}$ are the branch resistance and reactance. Because the network topology may change under NTR, (1) is imposed using a branch status variable $J_{ij,t}$ and a branch specific parameter $M_{ij}$ calculated from the admissible squared voltage ranges of the terminal buses. Specifically, the implementation uses the maximum possible terminal squared voltage difference plus a $10^{-3}$ numerical margin, avoiding an arbitrarily large big-$M$ value:

$$-M_{ij}\left(1 - J_{ij,t}\right) \leq v_{i,t} - v_{j,t} - 2\left(r_{ij}P_{ij,t} + x_{ij}Q_{ij,t}\right) + \left(r_{ij}^2 + x_{ij}^2\right)\ell_{ij,t} \leq M_{ij}\left(1 - J_{ij,t}\right) \tag{2}$$

The nonlinear branch power relation is represented through the second order cone relaxation

$$P_{ij,t}^2 + Q_{ij,t}^2 \leq v_{i,t}\ell_{ij,t} \tag{3}$$

Nodal voltage and branch current limits are enforced as

$$\left(V_i^{min}\right)^2 \leq v_{i,t} \leq (V_i^{max})^2 \tag{4}$$

$$0 \leq \ell_{ij,t} \leq \left(I_{ij}^{max}\right)^2 J_{ij,t} \tag{5}$$

The active and reactive nodal power balances include substation supply, PV generation, dispatchable generation, feeder BESS operation, background demand, data center demand, and branch losses. The active power balance at node $n$is written compactly as

$$\sum_{(i,n)\in E}\left(P_{in,t} - r_{in}\ell_{in,t}\right) - \sum_{(n,j)\in E} P_{nj,t} + P_{n,t}^{sub} + P_{n,t}^{PV} + P_{n,t}^{gen} + P_{n,t}^{BESS} = P_{n,t}^{BG} + P_{n,t}^{DC} \tag{6}$$

with an analogous reactive power balance. At each substation $s$, active and reactive power injections are bounded by $0 \leq P_{s,t}^{sub} \leq P_s^{sub,max}$ and $-Q_s^{sub,max} \leq Q_{s,t}^{sub} \leq Q_s^{sub,max}$, respectively. The formulation therefore retains the AC branch flow relationships while allowing topology, voltage control, DER dispatch, and data center operation to be optimized simultaneously. This structure follows the MISOCP network representation used in the underlying CVR and NTR formulation.

## *2.2 Background Load and Data Center Demand*

The conventional feeder demand is represented using a ZIP model. The active background demand at node $n$ and hour $t$ is

$$P_{n,t}^{BG} = \lambda^{BG} P_n^0 L_t^{BG} \left[K_Z^P \frac{v_{n,t}}{V_n^{rated^2}} + K_I^P \tilde{V}_{n,t} + K_P^P\right] \tag{7}$$

where $P_n^0$ is the nominal nodal demand, $L_t^{BG}$ is the hourly background load multiplier, and $\lambda^{BG}$ is the background loading scale used in the sensitivity analysis. The voltage magnitude term associated with the constant current component is approximated as

$$\tilde{V}_{n,t} = \frac{1}{2}\left(\frac{v_{n,t}}{V_n^{rated^2}} + 1\right) \tag{8}$$

The reactive demand follows the corresponding ZIP formulation using $K_Z^Q$, $K_I^Q$, and $K_P^Q$. Capacitor injections are modeled separately and are therefore not scaled when background demand is varied.

The AI data center is modeled independently from the ZIP demand. Let $C^{DC}$ denote the peak IT capacity to be determined by the optimization and let $L_t^{DC}$ denote the normalized workload profile. The baseline IT demand is

$$\bar{P}_t^{IT} = C^{DC} L_t^{DC} \tag{9}$$

Because the unshifted workload profile is peak normalized, $\max_{t\in T} L_t^{DC} = 1$. Therefore, $C^{DC}$ defines the installed IT power capacity referenced to the peak of that baseline profile. After workload scheduling is enabled, the realized hourly demand $P_t^{IT}$ remains bounded by $C^{DC}$ but need not attain it.

The total facility demand includes cooling, power conversion, and other supporting loads through power usage effectiveness:

$$P_t^{FAC} = PUE\, P_t^{IT} \tag{10}$$

Unlike (7), (9) and (10) contain no voltage dependent term. Consequently, reducing feeder voltage does not directly reduce AI computing demand.

## *2.3 Data Center Workload Flexibility*

All scheduling intervals are 1 h, so the workload conservation equations may be written directly in terms of hourly average power. Let $\varphi$ denote the fraction of arriving workload that is flexible and let $H$ denote the maximum allowable deferral duration.

The nonnegative variable $X_{\tau,t}$ represents the average IT power allocation during hour $t$ associated with flexible workload arriving at hour $\tau$, with $X_{\tau,t} \geq 0$. Work may only be executed between its arrival time and the end of the allowable scheduling window:

$$X_{\tau,t} = 0,\ t \notin \{\tau, \dots, \min(\tau + H, 24)\} \tag{11}$$

All flexible computational work must eventually be executed:

$$\sum_{t=\tau}^{\min(\tau+H,24)} X_{\tau,t} = \varphi\, \bar{P}_{\tau}^{IT} \tag{12}$$

For workload arriving near the end of the scheduling horizon, the allowable deferral window is truncated at 24 hours, preventing computational work from being carried beyond the modeled day. This same day completion requirement is imposed because each 24-hour profile is evaluated as an independent representative operating scenario, thereby avoiding unmodeled workload carryover between consecutive days. The actual IT power at hour $t$ is therefore

$$P_t^{IT} = (1 - \varphi)\, \bar{P}_t^{IT} + \sum_{\tau \in T} X_{\tau,t} \tag{13}$$

To ensure that workload scheduling does not exceed the installed computing capacity,

$$0 \leq P_t^{IT} \leq C^{DC} \tag{14}$$

Equations (11) - (13) conserve the total computational work. The model therefore shifts flexible computing demand rather than curtailing it. This distinction is important because any increase in interconnection capacity results from temporal coordination rather than a reduction in the required computing service. The implemented code follows this forward deferral and workload conservation structure directly.

## *2.4 Dedicated UPS Battery Model*

The data center UPS is represented separately from the BESS resources already installed within the distribution network. Its purpose is primarily to provide backup energy, although part of its capacity may support grid operation when the emergency reserve requirement remains satisfied.

Let $P_t^{ch}$, $P_t^{dch}$, and $E_t^{UPS}$ represent UPS charging power, discharging power, and stored energy. Binary variables $u_t^{ch}$ and $u_t^{dch}$ prevent simultaneous charging and discharging:

$$u_t^{ch} + u_t^{dch} \leq 1 \tag{15}$$

$$0 \leq P_t^{ch} \leq P_{\max}^{UPS,flex} u_t^{ch} \tag{16}$$

$$0 \leq P_t^{dch} \leq P_{\max}^{UPS,flex} u_t^{dch} \tag{17}$$

The power limit $P_{\max}^{UPS,flex}$ represents the portion of UPS power capability made available for normal grid support operation and does not represent the total emergency discharge rating of the UPS system. The emergency UPS power capability is assumed to be sized to support the specified critical facility demand, while the reserve constraint below ensures that sufficient stored energy is retained for the required backup duration.

The stored energy evolves according to

$$E_t^{UPS} = E_{t-1}^{UPS} + \Delta t \left( \eta^{ch} P_t^{ch} - \frac{P_t^{dch}}{\eta^{dch}} \right) \tag{18}$$

The initial UPS energy is fixed as $E_0^{UPS} = SOC^{init} E_{\max}^{UPS}$. The energy capacity is constrained by

$$0 \le E_t^{UPS} \le E_{\max}^{UPS} \tag{19}$$

To prevent grid support operation from encroaching on the UPS backup function, a nominal emergency energy reserve is maintained whenever UPS flexibility is enabled. The reserve is represented as an energy-based planning requirement equivalent to the specified fraction of peak facility demand over the required reserve duration. Emergency mode conversion losses and detailed outage discharge dynamics are outside the scope of the interconnection model; the charge and discharge efficiencies in (18) apply to normal grid support operation. Unlike unrestricted grid storage, the UPS must continuously retain sufficient energy for emergency operation. The minimum reserve is defined as

$$E_t^{UPS} \ge h^{res} \rho^{crit} PUE\ C^{DC} \tag{20}$$

where $h^{res}$ is the required reserve duration and $\rho^{crit}$ is the fraction of peak facility demand that must remain supported during an emergency.

A cyclic terminal condition is imposed:

$$E_{24}^{UPS} = SOC^{init} E_{\max}^{UPS} \tag{21}$$

Together with the reserve requirement in (20), the fixed boundary SOC imposes the capacity bound $C^{DC} \le \frac{SOC^{init} E_{\max}^{UPS}}{(h^{res} \rho^{crit} PUE)}$ whenever UPS reserve is enforced. Daily discharge throughput is limited by

$$\sum_t P_t^{dch} \Delta t \le N^{cyc} E_{\max}^{UPS} \tag{22}$$

The throughput limit is expressed on a delivered discharge energy basis; $N^{cyc} = 1$ therefore represents one rated energy equivalent of delivered UPS discharge over the scheduling horizon. The active power drawn by the data center from the distribution network is therefore

$$P_{n,t}^{DC} = P_t^{FAC} + P_t^{ch} - P_t^{dch} \tag{23}$$

$$P_{n,t}^{DC} \ge 0 \tag{24}$$

Thus, UPS can reduce grid demand during critical hours but cannot export power from the data center to the network. Data center demand is zero at all other buses. The emergency reserve and throughput limits also prevent the backup system from being treated as unrestricted grid storage. The reserve requirement is defined relative to peak facility demand rather than instantaneous scheduled demand and therefore maintains a fixed nominal backup energy reservation throughout the scheduling horizon.

### *2.5 Grid Side Flexibility*

CVR is represented through continuous control of the voltage at each substation bus. When CVR is disabled, the corresponding feeder head voltage is fixed at its nominal operating value. When CVR is enabled, the squared voltage may be reduced within the allowable CVR and network voltage limits:

$$\left(V_s^{CVR,\min}\right)^2 \le v_{s,t} \le \left(V_s^{base}\right)^2,\ s \in S,\ t \in T \tag{25}$$

The voltage changes propagate through the branch flow equations and modify the voltage dependent ZIP background demand in (7). The AI data center load remains voltage independent, so CVR creates interconnection headroom only by reducing surrounding feeder demand. When CVR is disabled, the substation voltage is fixed at its nominal value. CVR is modeled as continuous substation voltage control within the admissible steady state range; discrete regulator taps and switching transients are not represented.

NTR modifies only a predefined set of controllable branches, while all other branch statuses remain fixed. The binary variable $J_{ij,t}$ indicates whether branch $(i,j)$ is energized. For a network with $|N|$ buses and $|S|$ substations, the required number of energized branches is

$$\sum_{(i,j)\in E} J_{ij,t} = |N| - |S|,\ t \in T \tag{26}$$

The branch count condition in (26) is combined with a single commodity artificial flow formulation to enforce connectivity and radiality [37]. Artificial flow is allowed only on energized branches, with each non substation bus consuming one unit and substations supplying the total required flow. Switching activity is measured relative to the initial branch status in the first period and the previous period, thereafter, using the standard binary linearization of the absolute difference in (27).

$$\sum_{t\in T} \sum_{(i,j)\in E^{sw}} a_{ij,t} \le A^{\max} \tag{27}$$

Here, $a_{ij,t}$ indicates a branch status change and $A^{\max}$ is the allowable switching budget. No terminal topology restoration is enforced, allowing the final configuration to remain in service beyond the scheduling horizon. The switching budget therefore limits excessive topology changes.

When NTR is disabled, all branches remain in the base radial configuration. When enabled, only designated controllable branches may switch, allowing the GRID and FULL cases to jointly optimize feeder topology, voltage, and other operating variables.

### *2.6 Existing Distributed Energy Resources*

The distribution network includes PV, dispatchable generation, and feeder BESS, which are kept unchanged across all cases so that capacity differences reflect only the evaluated flexibility mechanisms. PV curtailment, generator power limits, and standard BESS energy balance, SOC, charge/discharge, apparent power, and cyclic terminal energy constraints are enforced following [37]. The cyclic condition prevents artificial capacity gains from end-of-horizon BESS depletion.

### *2.7 Interconnection Capacity Objective and Solution Procedure*

The primary objective is to maximize the installed IT capacity while satisfying all network, workload, UPS, voltage, thermal, radiality, and resource constraints:

$$\max\ C^{DC} \tag{28}$$

Because the workload profiles are peak normalized, the optimized capacity is directly comparable across cases.

A second stage fixes the Stage 1 capacity and discrete decisions, then minimizes feeder active energy losses:

$$\min \sum_{t\in T} \sum_{(i,j)\in E} r_{ij}\, \ell_{ij,t} \Delta t \tag{29}$$

This stage refines only the continuous operating variables without changing the accepted interconnection capacity.

Because (3) uses an SOCP relaxation, solution tightness is verified using the normalized branch relaxation slack in (30). The maximum normalized slack over all energized branches and time periods must remain below $10^{-3}$. If necessary, Stage 2 is repeated with tighter numerical tolerances. No reported case required a reduction in interconnection capacity after this refinement.

$$\delta_{ij,t} = \frac{max(0, v_{i,t}\ell_{ij,t} - P_{ij,t}^2 - Q_{ij,t}^2)}{(S_{ij}^{max})^2} \tag{30}$$

### *2.8 Operating Configurations*

Four primary operating configurations are considered to isolate the benefits of grid side flexibility, data center flexibility, and their coordinated operation. The configuration logic implemented in the model is summarized in Table 1.

Table 1. Primary operating configurations for AI data center interconnection assessment.

| *Case* | *CVR* | *NTR* | Workload Scheduling | UPS Flexibility |
|---|---|---|---|---|
| BASE | No | No | No | No |
| GRID | Yes | Yes | No | No |
| DCFLEX | No | No | Yes | Yes |
| FULL | Yes | Yes | Yes | Yes |

BASE represents the conventional interconnection case with fixed feeder voltage, fixed topology, rigid workload, and no use of the dedicated UPS for network support. GRID activates CVR and NTR while retaining a rigid data center. DCFLEX retains the base network configuration but enables workload scheduling and reserve constrained UPS operation. FULL coordinates all four flexibility mechanisms.

Additional CVR, NTR, WORKLOAD, and UPS configurations are evaluated in the mechanism decomposition analysis to separate the individual contribution of each flexibility resource. These auxiliary configurations use the same network, load, DER, and data center assumptions as the four primary cases and differ only in the enabled control mechanisms.

In BASE and GRID, the dedicated UPS is not dispatched for network support, and its emergency function is assumed unaffected by the operational optimization. In DCFLEX and FULL, UPS charging and discharging are enabled subject to the grid-support power limit, energy balance, throughput, and nominal emergency reserve constraints.

## 3. Case Study and Experimental Design

### *3.1 AI Data Center Adapted IEEE 123 Bus System*

The proposed framework is evaluated on a balanced network derived from the IEEE 123-node feeder [35] and adapted from [37]. The model contains 116 buses, 120 candidate branches, three substations, denoted substations 1–3 at buses 48, 65, and 76, respectively, six PV units, six feeder BESS units, and two dispatchable generators over a 24 h horizon. Eight branches (53, 59, 63, 111, 112, 116, 117, and 118) are controlled under a four-action switching budget, limiting daily reconfiguration to at most two open close exchanges; the reported NTR capacities are therefore conditional on this policy. The 113-branch base radial topology has controllable branches 53, 59, and 63 initially closed and 111, 112, 116, 117, and 118 open, while branches 119 and 120 remain fixed open. The underlying network and DER locations follow [37], while the data center connection, source limits, reinforced thermal ratings, and flexibility models are introduced here.

The adapted network, data center connection, DER locations, and controllable branches are shown in Figure 1. The feeder voltage is increased from 4.16 kV to 13.8 kV on a 100 MVA base, with impedances rescaled accordingly. Branch ampacity limits of 800 A and 1000 A correspond to approximately 19.12 MVA and 23.90 MVA, and the three substation active power limits are 20, 16, and 20 MW. The primary data center is connected at bus 60 in the substation 2 area, where BASE operation reaches the feeder minimum voltage limit before the 16 MW source limit, allowing NTR to be evaluated for accessing headroom elsewhere in the network. Nominal background demand is 3.49 MW and 1.92 MVAr, with a 24% constant impedance, 36% constant current, and 40% constant power ZIP composition applied to both active and reactive demand. Bus voltages are limited to 0.95-1.05 p.u.; substation voltage is nominally 1.05 p.u. and may be reduced to 0.95 p.u. under CVR. The resulting system is a physically scaled synthetic test network rather than a specific utility feeder.

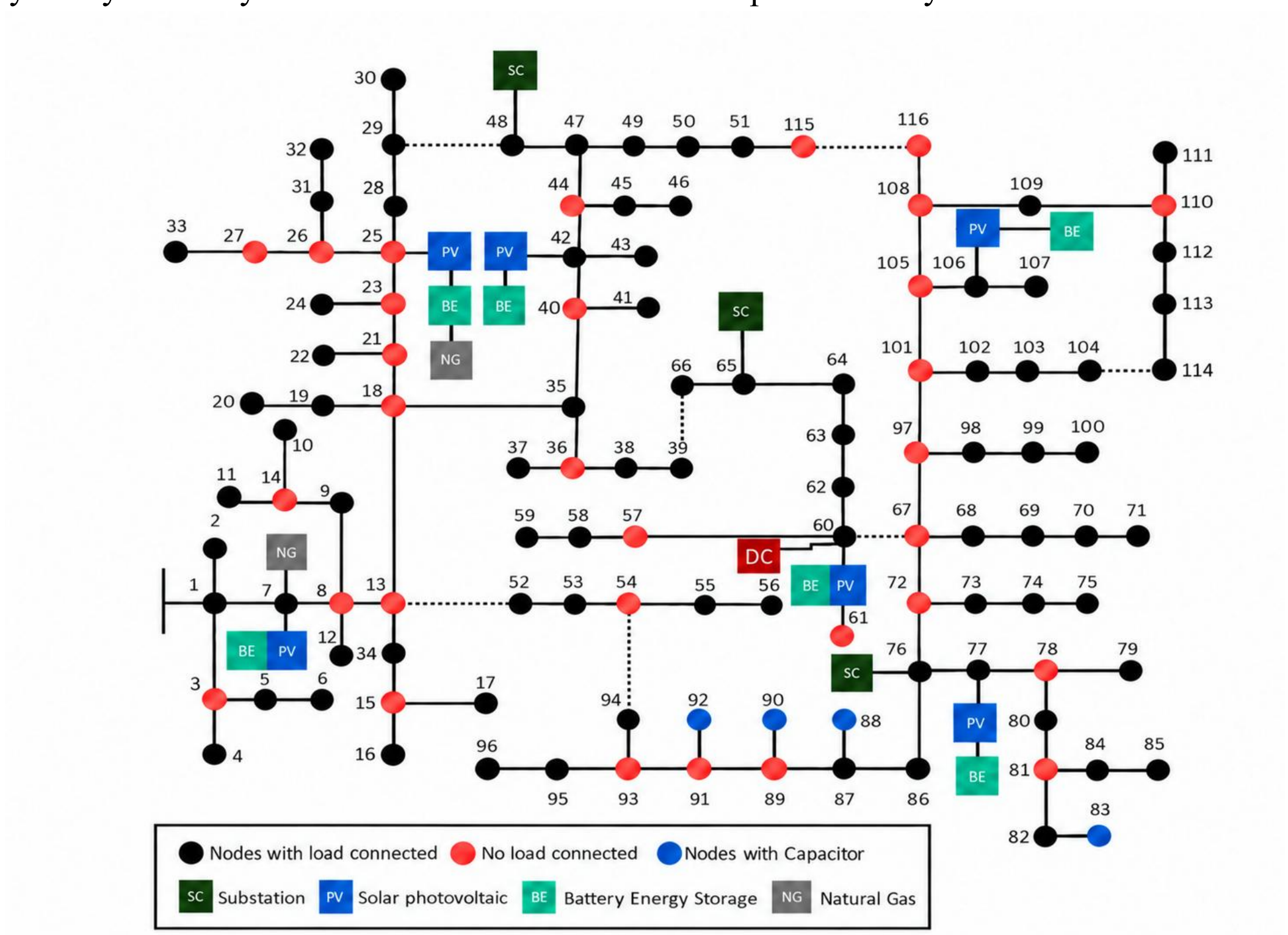

Figure 1 AI load adapted IEEE 123-node distribution system with DERs and data center connection.

Bus 60 is used as the primary data center connection point because it lies within the area associated with the lower rated 16 MW source and exhibits a comparatively constrained base topology operating condition. Buses 35 and 90 are additionally evaluated to assess the sensitivity of interconnection capacity to connection location. The primary bus 60 case therefore represents a constrained siting condition in which relocation to a stronger connection point is assumed unavailable or undesirable.

### *3.2 AI Data Center Workload Profiles*

Three representative data center workload archetypes are considered: Training, Inference, and Mixed. Annual synthetic data center load series are generated using the Lawrence Berkeley National Laboratory Data Center and Industrial Electrical Load Shape Maker [8]. The tool allows data center usage characteristics and temporal behavior to be configured and provides a reproducible means of generating representative load profiles when measured AI data center traces are unavailable.

For each workload type, a representative weekday is selected from the corresponding annual series and reduced to a 24-hour profile. Each profile is normalized by its own daily maximum such that its peak value is unity. Consequently, the peak normalization places all workload archetypes on a commonly installed IT capacity basis. Figure 2 compares the resulting normalized workload profiles. The selected representative days are May 13 for Training, December 19 for Inference, and August 27 for Mixed operation. These dates are used only to identify representative daily profiles and are not intended to represent chronological coincidence with the feeder background load.

As shown in Figure 2, the Training profile remains close to its peak throughout the day and has a load factor of 98.68%, representing sustained high utilization. The Inference profile exhibits greater temporal variation than the Training profile while maintaining relatively high utilization, resulting in a load factor of 92.28%. The Mixed profile combines characteristics of both cases and has an intermediate load factor of 95.38%. These distinct temporal patterns enable the analysis to evaluate how workload characteristics influence interconnection capacity and the value of operational flexibility.

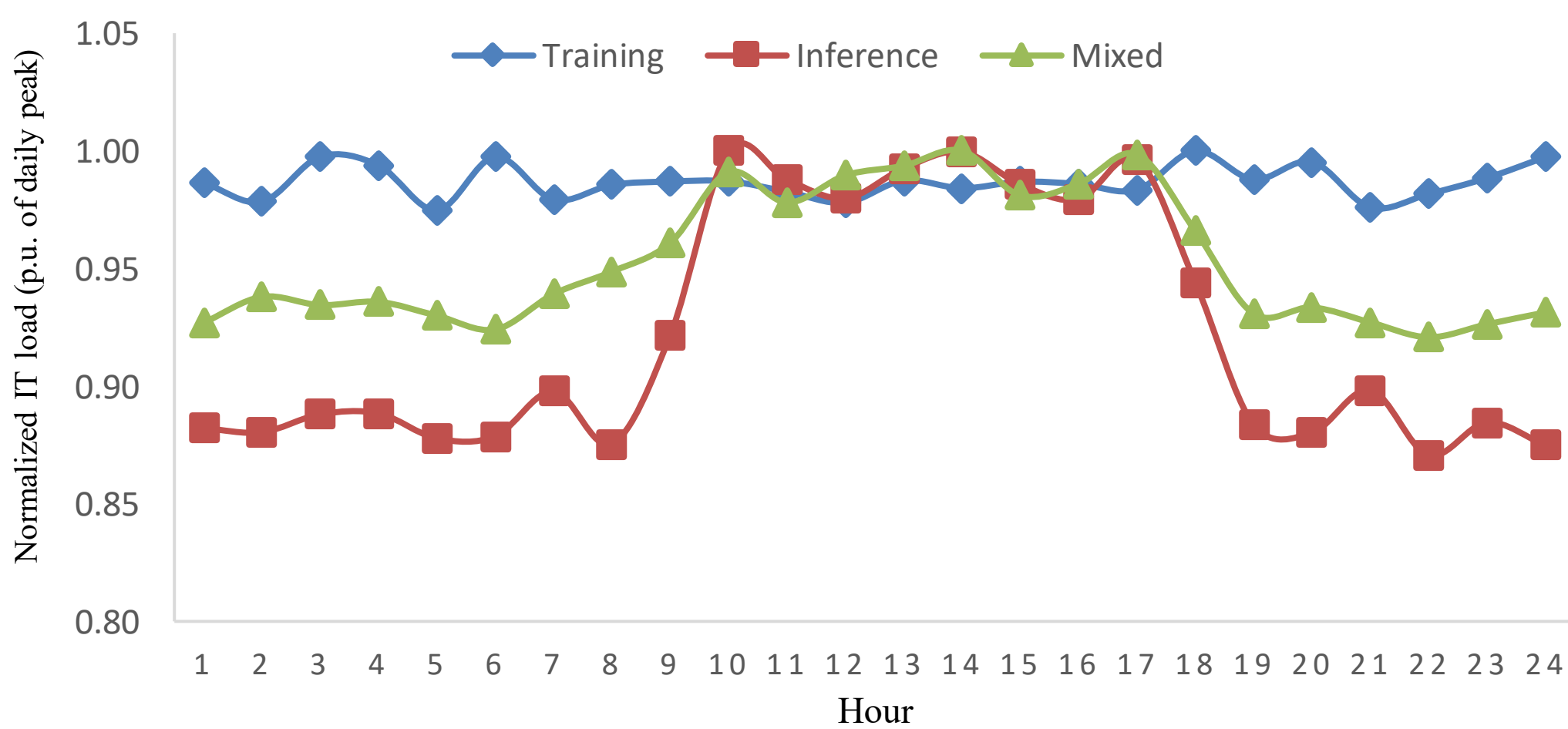

Figure 2 Representative 24-hour AI data center workload profiles.

The selected profiles are intended to represent different AI computing behaviors rather than reproduce a specific facility operating on a particular calendar day. Therefore, the representative dates used during profile extraction do not imply chronological coincidence with the feeder background load profile. All network, data center, and operational parameters are otherwise maintained consistently across the three workload cases so that differences in the results can be attributed primarily to workload shape.

### *3.3 Data Center and UPS Parameters*

A baseline PUE of 1.20 and unity power factor are assumed. In flexible cases, 5% of the IT workload may be deferred within 3 h. The UPS has a 6 MWh energy capacity, with up to 3 MW of charging or discharging power available for grid-support operation, 0.95 charging and discharging efficiencies, and 80% initial and terminal SOC. It maintains a 0.5 h emergency reserve equal to 50% of peak facility demand and is limited to one rated-energy equivalent of delivered discharge per day.

The workload flexibility, UPS capacity, grid support power, and reserve assumptions are treated as representative baseline values and are examined through sensitivity analysis. UPS ratings remain fixed as interconnection capacity varies, so the sensitivities evaluate the effect of a given backup resource rather than jointly optimizing UPS size and data center capacity.

### *3.4 Experimental Design*

The primary analysis considers the four operating configurations defined in Table 1 for the Training, Inference, and Mixed workload profiles, resulting in 12 principal interconnection cases. BASE represents the reference condition without additional flexibility, GRID activates CVR and NTR, DCFLEX enables workload scheduling and UPS operation, and FULL coordinates both grid and data center flexibility. Additional analyses isolate the individual contributions of CVR, NTR, workload scheduling, and UPS operation and evaluate the sensitivity of interconnection capacity to connection location, PUE, flexible workload fraction, deferral duration, UPS energy and reserve requirements, and background feeder loading. The specific ranges considered in each analysis are presented with the corresponding results. The interaction between grid side and data center side flexibility is quantified as

$$\Delta C^{int} = C_{\mathrm{FULL}} - C_{\mathrm{GRID}} - C_{\mathrm{DCFLEX}} + C_{\mathrm{BASE}}$$

A positive value indicates a positive interaction in absolute MW under the adopted additive capacity metric, whereas a negative value indicates overlapping effects. This metric does not imply that the proportional effectiveness of one flexibility source necessarily increases in the presence of the other. Because the operating configurations also differ in the reliability constraints associated with enabled UPS grid support operation, $\Delta C^{int}$ is used here only as a descriptive difference in differences measure across the defined cases and is not interpreted as a formal supermodularity or complementarity metric. The relative interaction is reported as $\frac{100\Delta C^{int}}{C_{\mathrm{BASE}}}$. A separate adverse operating condition is also evaluated for the Mixed workload at bus 60 to test whether the observed flexibility benefits remain under a more restrictive combination of feeder loading, PV availability, and temporal load coincidence. The adverse operating condition combines 110% background feeder loading with zero PV availability, used as a conservative proxy for a low PV condition because only one PV day is available in the source data. The Mixed workload profile is shifted by 5 h so that its original peak at hour 14 coincides with the feeder peak at hour 19.

### *3.5 Computational Settings and Numerical Screening*

The model is implemented in Pyomo and solved with Gurobi [36]. Stage 1 maximizes interconnection capacity with a 0.1% MIP gap target and 500 s time limit. Stage 2 fixes the accepted capacity and discrete decisions, then solves a continuous SOCP loss minimization with the same time limit.

After each solution, numerical feasibility and SOCP relaxation quality are checked, with the maximum normalized relaxation slack required to remain below $10^{-3}$. If necessary, Stage 2 is repeated at the same capacity using tighter tolerances. No reported case required capacity reduction. All 12 principal cases and the adverse Mixed FULL case are independently validated using balanced AC power flow replay.

## 4. Results and Discussion

### *4.1 Coordinated Flexibility and AI Data Center Interconnection Capacity*

Figure 3 compares the maximum feasible IT capacity under the four operating configurations ordered by delivered capacity. BASE accommodates 10.48 MW, 10.61 MW, and 10.58 MW for Training, Inference, and Mixed workloads, respectively. GRID provides the largest increase, raising capacity to 14.93-15.38 MW, corresponding to gains of about 42.5-45.0%. DCFLEX provides smaller gains, increasing capacity to 10.70-11.18 MW. FULL achieves the highest capacities of 15.29 MW, 16.00 MW, and 15.88 MW, corresponding to gains of about 45.9-50.8% over BASE. The 16 MW Inference FULL case reaches the UPS reserve ceiling. Overall, grid side flexibility provides most of the capacity gain, while data center flexibility provides additional headroom.

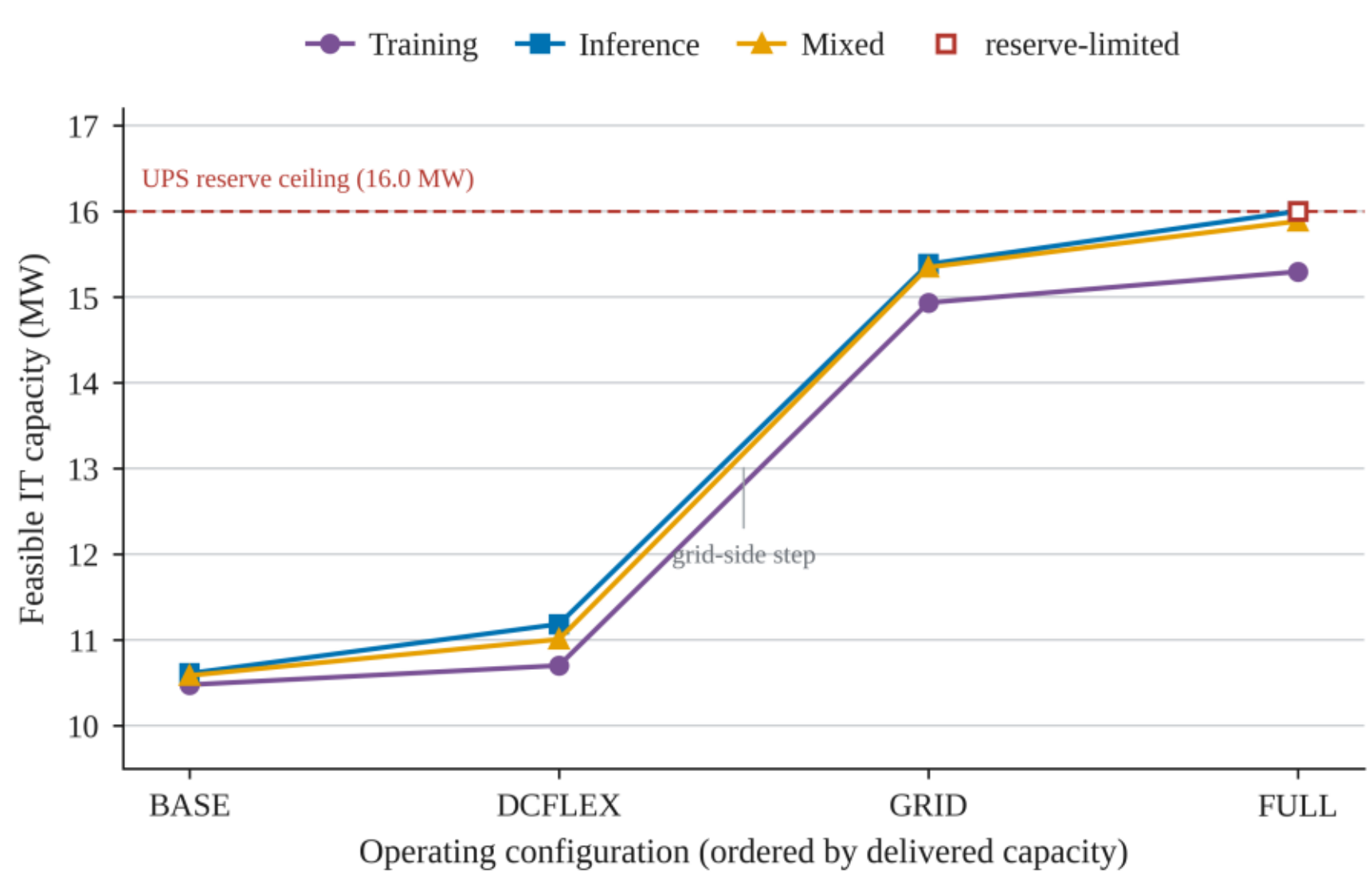

Figure 3 Feasible IT capacity across operating configurations and workload archetypes.

The retained Mixed FULL topology illustrates the transfer mechanism underlying the NTR benefit. At hour 1, branch 59 (60-62) opens and branch 112 (60-67) closes, transferring bus 60 and the associated buses 52–61 from the substation 2 supply area toward substation 3. At hour 12, branch 53 (54-57) opens and branch 111 (13-52) closes, producing a second feeder boundary adjustment in the retained Stage 1 solution. No additional topology changes occur during the remaining hours.

The 16.00 MW capacity obtained for the Inference FULL case reaches the ceiling imposed by the adopted nominal UPS reserve requirement and the 80% initial and terminal SOC. Under the reserve formulation

$$E_t^{UPS} \geq h^{res} \cdot \rho^{crit} \cdot PUE \cdot C^{DC}$$

With the fixed initial and terminal stored energy $E_0^{UPS} = E_{24}^{UPS} = 0.8 \times 6 = 4.8MWh$, $h^{res} = 0.5h$, $\rho^{crit} = 0.5$, $PUE = 1.2$, the corresponding capacity ceiling is

$$C^{DC} \leq \frac{4.8}{0.5 \times 0.5 \times 1.2} = 16\ MW$$

Accordingly, the Inference FULL case is reserve limited under the adopted UPS assumptions rather than by the unrestricted network limit. In BASE, the binding condition is the local voltage limit, not the 16 MW substation 2 rating; in the Inference BASE case, bus 52 reaches 0.950 p.u. while substation 2 peaks at about 14.30 MW.

With NTR, the feeder around bus 60 is reconfigured to shift part of the data center and surrounding load toward substation 3, relieving the local voltage constraint. In the Mixed GRID case, substation 3 reaches about 20 MW, while the four-action switching budget is fully used, indicating that the limiting condition shifts to substation 3 capacity and the permitted reconfiguration budget. FULL coordination then adds further headroom through the combined data center flexibility mechanisms.

### *4.2 Contribution of Individual Flexibility Mechanisms*

Figure 4 shows that NTR provides nearly all of the grid side capacity gain at bus 60. For the Mixed workload, CVR provides essentially no gain, while NTR contributes 4.762 MW above BASE, raising capacity from 10.585 MW to 15.347 MW. On the data center side, workload scheduling and UPS operation contribute 0.337 MW and 0.396 MW individually, while their combined DCFLEX gain is 0.421 MW, indicating substantial overlap between these mechanisms. FULL provides the largest gain of 5.299 MW above BASE.

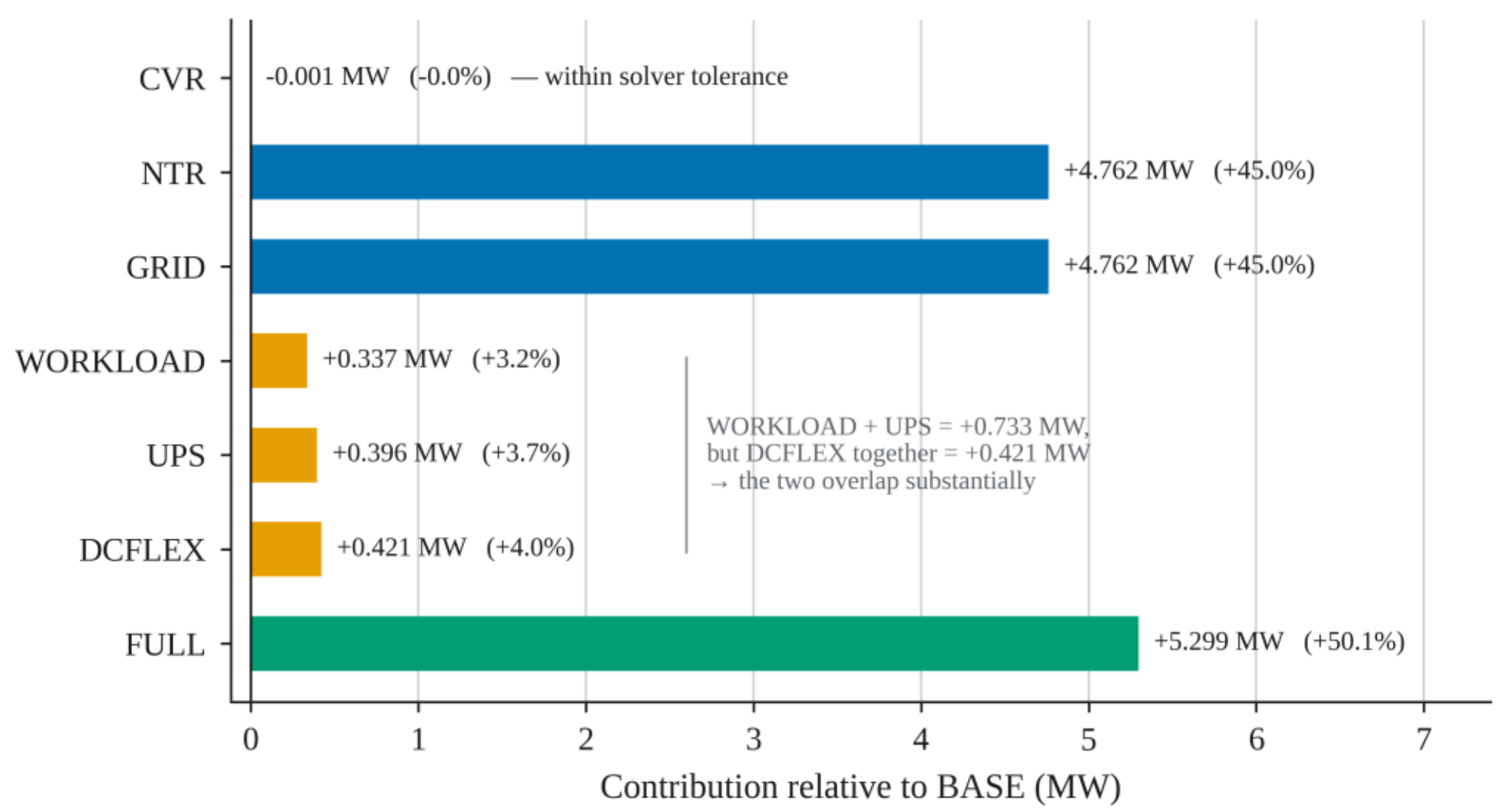

Figure 4 Contribution of individual grid side and data center side flexibility mechanisms to interconnection capacity for the Mixed workload at bus 60

This overlap within data center flexibility does not preclude a positive interaction between aggregate data center flexibility and grid side flexibility. The interaction metric $\Delta C^{int}$ is 0.137 MW, 0.043 MW, and 0.115 MW for Training, Inference, and Mixed workloads, corresponding to 1.31%, 0.41%, and 1.09% of BASE capacity, respectively. These positive values indicate a modest additional absolute benefit from coordinated flexibility. For Inference, however, FULL reaches the 16 MW UPS reserve ceiling while GRID and DCFLEX remain below it; therefore, the corresponding $\Delta C^{int}$ is reserve ceiling affected and should not be compared directly in magnitude with the Training and Mixed cases. Accordingly, $\Delta C^{int}$ is used only as a descriptive difference in differences metric and not as a formal measure of complementarity.

Table 2 Interaction between grid and data center flexibility

| ***Workload*** | $\Delta C^{int}$ ***(MW)*** | $\Delta C^{int}$ ***relative to BASE*** | ***Interpretation*** |
|---|---|---|---|
| Training | 0.137 | 1.31% | Positive descriptive interaction |
| Inference | 0.043 | 0.41% | Positive; reserve-limited |
| Mixed | 0.115 | 1.09% | Positive descriptive interaction |

### *4.3 Location sensitivity*

Figure 5 examines the effect of connection location for the Mixed workload. At bus 35, BASE and GRID both reach about 15.71 MW, while DCFLEX and FULL reach the 16 MW UPS reserve ceiling, indicating no measurable grid side capacity gain. Bus 60 shows the strongest improvement: GRID increases capacity from 10.58 MW to 15.35 MW, while FULL reaches 15.88 MW, about 50.1% above BASE, because NTR relieves the local voltage constraint. At bus 90, BASE and GRID are both about 14.83 MW, while DCFLEX and FULL reach about 15.38 MW. Overall, NTR is most valuable at locally constrained connection points, while data center flexibility provides smaller but more consistent gains across locations.

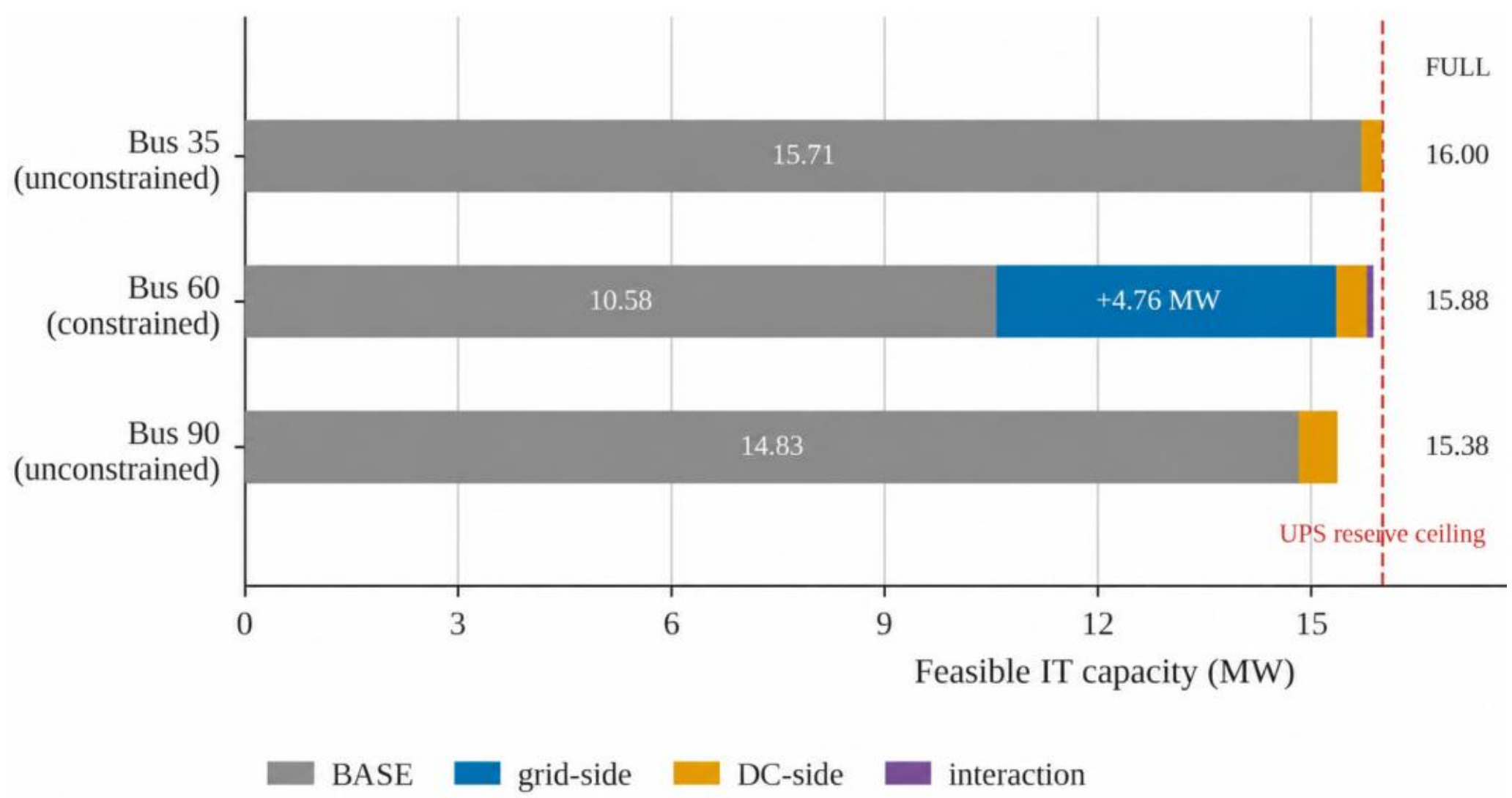

Figure 5 Effect of connection location on feasible IT capacity for the Mixed workload

### *4.4 Workload flexibility and deferral window*

Figure 6(a) shows that increasing the flexible workload fraction improves capacity only modestly under FULL operation. For Inference, capacity rises from 15.763 MW at 0% flexibility to the 16 MW UPS reserve ceiling at 5%, while larger fractions provide no additional increase. For Mixed, capacity increases from 15.739 MW to about 15.89 MW by 5–10% flexibility, with later differences within solver tolerance. Thus, most of the benefit is obtained with a small flexible fraction.

Figure 6(b) shows a similar trend for deferral duration. Inference capacity increases from 15.763 MW with no deferral to 16.000 MW at 3 h, again reaching the UPS reserve ceiling. For Mixed, capacity increases from 15.739 MW to about 15.89 MW by 2–3 h, with the small difference between these cases within numerical resolution. Overall, additional temporal flexibility provides diminishing gains as other network and UPS constraints become active.

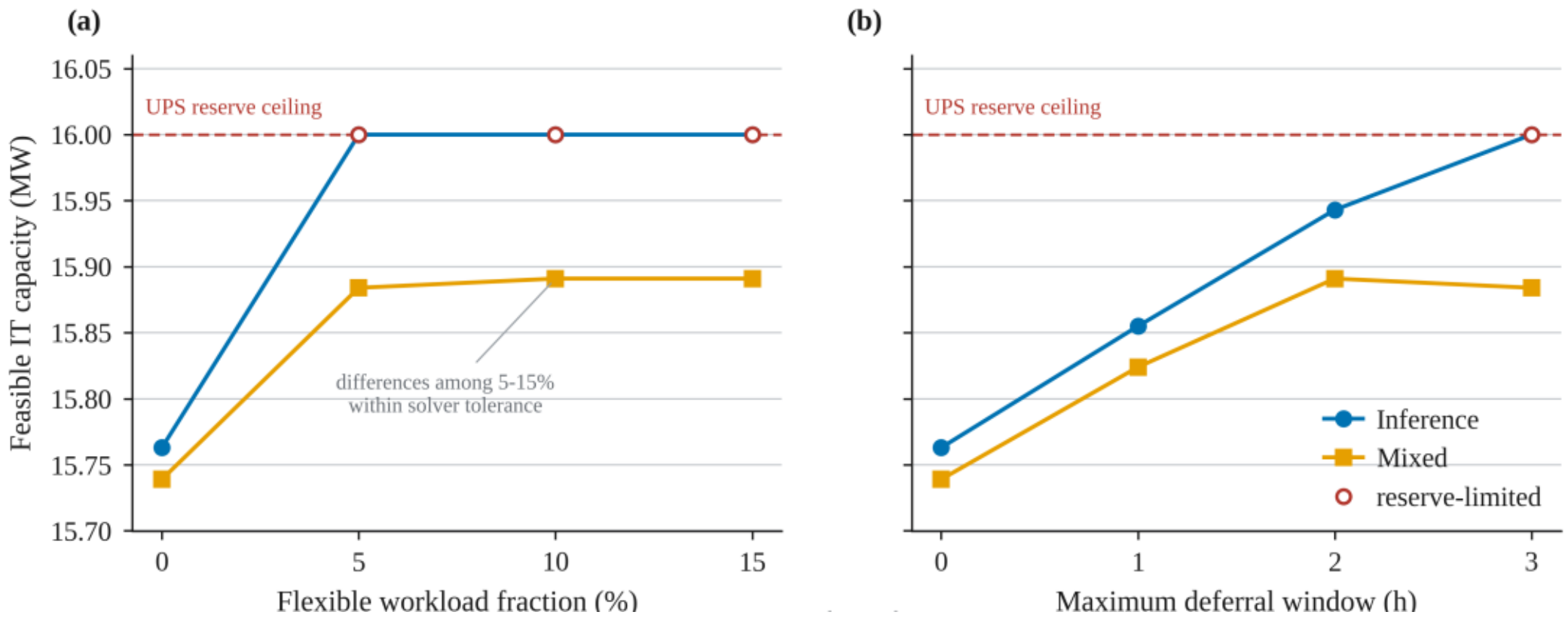

Figure 6 Effect of workload flexibility on interconnection capacity: (a) flexible workload fraction and (b) maximum deferral window

## *4.5 PUE, UPS energy, and reserve sensitivity*

Figure 7(a) shows that higher PUE reduces feasible IT capacity under Mixed FULL operation, from 17.328 MW at PUE 1.10 to 13.621 MW at 1.40, while facility-level capacity remains nearly constant at about 19.06 MW. Thus, changes in PUE primarily rescale the IT capacity corresponding to this approximately constant feasible facility demand.

Increasing UPS energy capacity also increases feasible IT capacity, but with diminishing returns: the Inference FULL case rises from 16.000 MW at 6 MWh to 16.207 MW at 8 MWh and 16.258 MW at 10 MWh. Reserve duration has a stronger effect because it directly determines the required emergency energy.

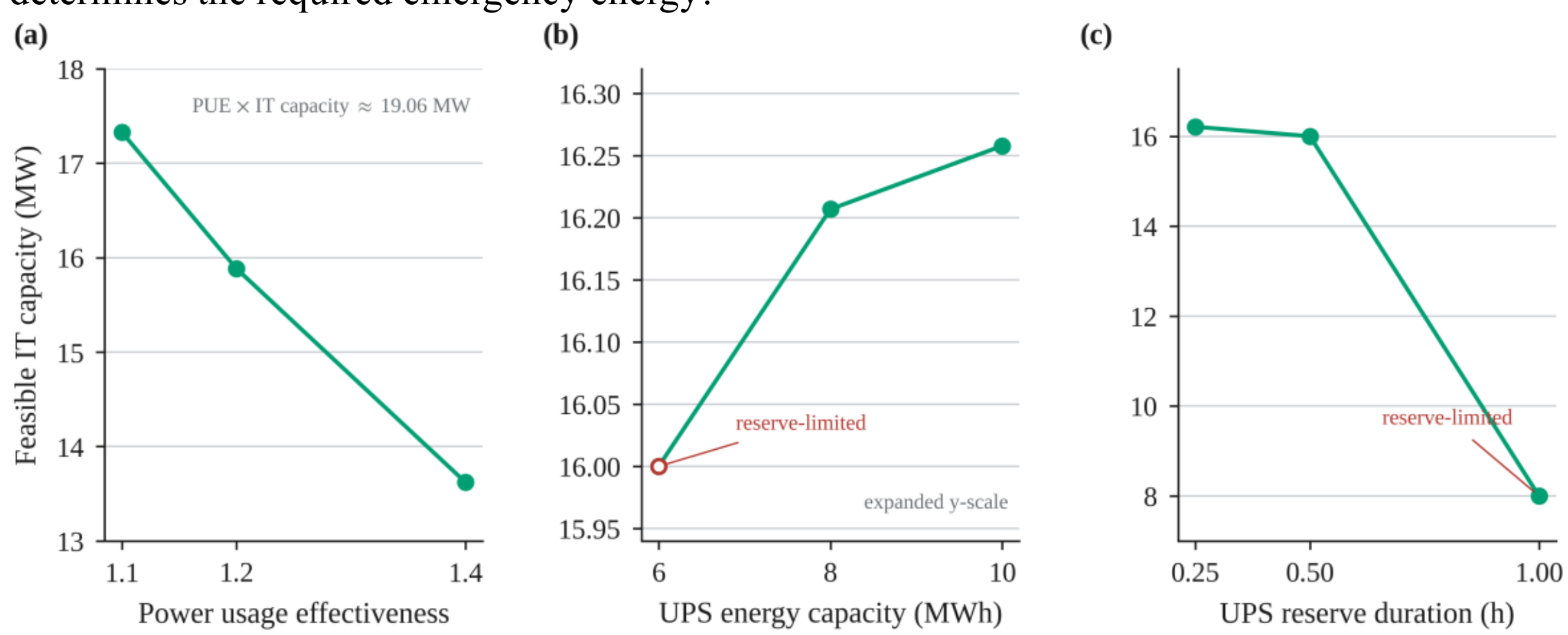

Figure 7 Effects of data center characteristics on interconnection capacity: (a) PUE, (b) UPS energy capacity, and (c) UPS emergency reserve duration.

At a 1.00 h reserve duration, the 8.00 MW result is set directly by the UPS reserve bound, whereas at 0.25 h the reserve ceiling is nonbinding and other system constraints determine the feasible capacity. This result reflects the backup-adequacy ceiling imposed by the 1 h reserve requirement and fixed boundary SOC; it should not be interpreted as a reduction caused by enabling UPS grid-support operation relative to a backup-compliant BASE case. Reducing reserve duration from 0.50 h to 0.25 h increases capacity to 16.211 MW, while increasing it to 1 h reduces capacity to 8 MW. Overall, UPS reserve assumptions can materially affect the capacity available for grid support.

### *4.6 Background Feeder Loading and Adverse Operating Condition*

Figure 8 shows that the benefit of coordinated flexibility remains stable as background feeder loading increases from 90% to 110%. FULL capacity decreases only from 15.974 MW to 15.808 MW, while the additional headroom over BASE decreases slightly from about 5.37 MW to 5.25 MW. Thus, moderate changes in background loading have little effect on the additional interconnection capacity provided by coordinated flexibility.

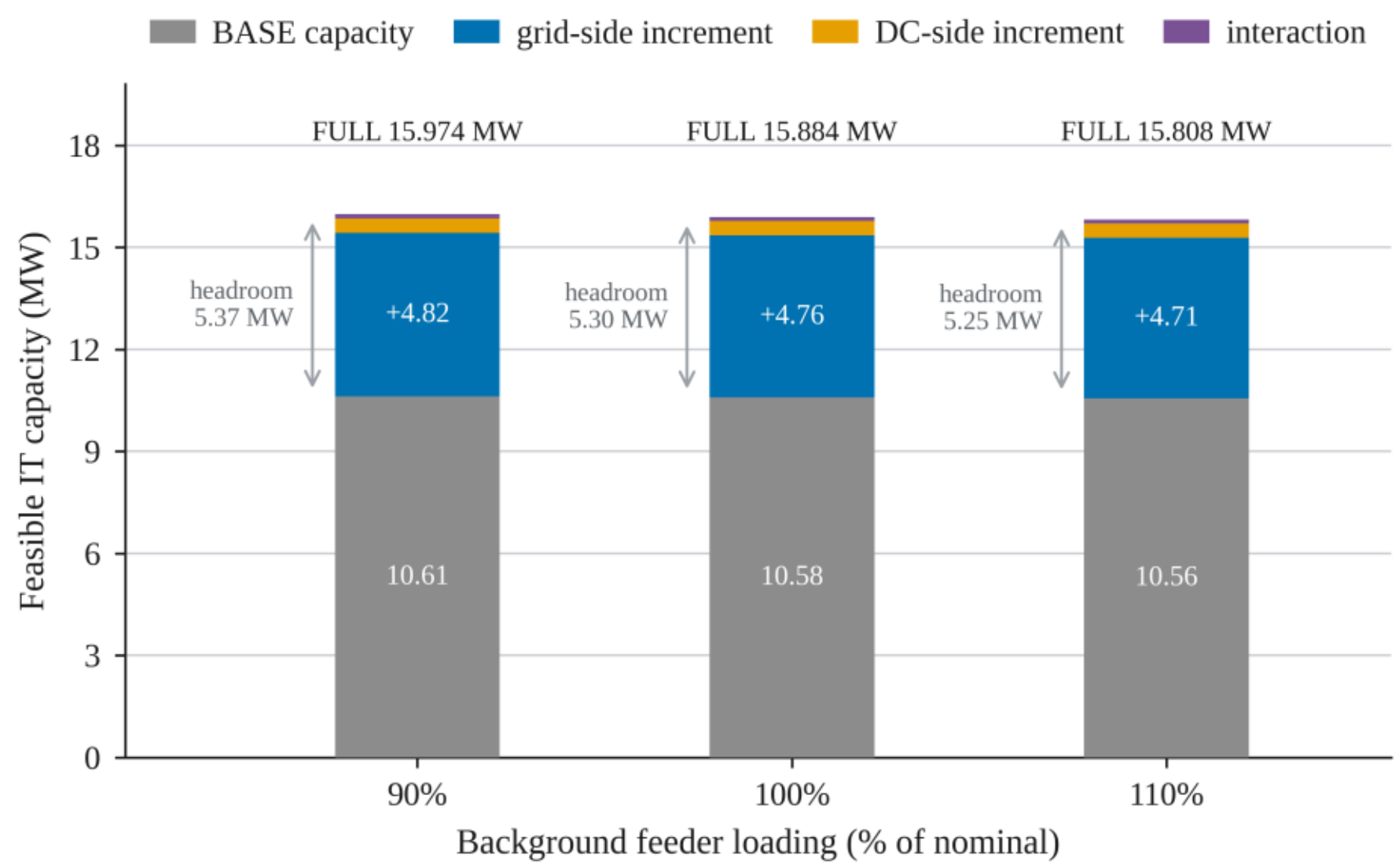

Figure 8 Decomposition of feasible IT capacity under varying background feeder loading.

As an additional robustness check, a combined adverse scenario is evaluated for the Mixed workload at bus 60 with 110% background loading, zero PV, and a 5 h workload shift. BASE, GRID, DCFLEX, and FULL accommodate 10.385, 14.718, 10.757, and 15.343 MW, respectively. FULL provides about 4.96 MW more capacity than BASE and 0.63 MW more than GRID. BASE and DCFLEX remain voltage limited, while GRID and FULL are constrained by substation 3 capacity and the switching budget. All cases satisfy the numerical screening criteria. Thus, the coordinated flexibility benefit persists under the combined stress condition of higher background loading, zero PV availability, and unfavorable workload timing. The reported values remain scenario-based interconnection limits rather than year-round firm ratings.

### *4.7 Numerical Screening and AC Power Flow Validation*

All 12 principal cases satisfy the numerical screening criteria. The maximum Stage 1 MIP gap is 0.0910%, within the 0.1% target, and the maximum normalized SOCP relaxation slack and power balance residuals remain below their acceptance limits. The Inference BASE case required tightened Stage 2 refinement but passed without reducing interconnection capacity and was also validated through AC power flow replay.

BASE and DCFLEX are generally limited by the feeder voltage constraint, while NTR shifts the limiting condition toward substation 3 capacity and the switching budget. In GRID and FULL, substation 3 reaches its 20 MW limit and all four permitted switching actions are used, confirming that the capacity gain is primarily driven by network reconfiguration and source redistribution.

Table 3 Numerical screening results for the principal MISOCP cases.

| *Metric* | *Maximum / minimum observed* | *Criterion* |
|---|---|---|
| Cases passing numerical audit | 12/12 | All |
| Maximum Stage 1 MIP gap | 0.091% | ≤ 0.1% |
| Maximum normalized SOCP relaxation slack | $3.70 \times 10^{-4}$ | $< 10^{-3}$ |
| Maximum active balance residual | $4.43 \times 10^{-7}$ | Near zero |
| Maximum reactive balance residual | $3.8 \times 10^{-7}$ | Near zero |
| Minimum nodal voltage | 0.95 p.u. | ≥ 0.95 p.u. |

The AC replay set comprised all 12 principal operating cases together with the adverse Mixed FULL case, totaling 13 cases and 312 hourly operating points. All 312 AC power flows converged successfully and satisfied the voltage, thermal, and source capacity limits. The maximum voltage magnitude deviation between the optimization and AC power flow solutions was $4.48 \times 10^{-6}$ p.u., confirming that the accepted MISOCP solutions closely represent the corresponding nonlinear AC operating points.

## 5. Conclusion

This study developed a coordinated flexibility framework for quantifying AI data center interconnection capacity in distribution networks. The formulation jointly represents utility side CVR and NTR with data center workload shifting and reserve constrained UPS operation within a two-stage MISOCP solution procedure. Across the Training, Inference, and Mixed workloads at the primary bus-60 connection point, BASE accommodates approximately 10.5-10.6 MW of IT capacity. Grid side flexibility increases this to approximately 14.9-15.4 MW, corresponding to gains of 42.5-45.0%, while data center flexibility alone provides smaller gains of approximately 2.1-5.4%. Coordinated FULL operation increases feasible IT capacity to approximately 15.3-16.0 MW, representing improvements of approximately 45.9–50.8% over BASE, with the 16.0 MW Inference result limited by the adopted UPS reserve requirement.

At the primary constrained bus-60 location, NTR is the dominant source of grid side capacity improvement because it redistributes feeder supply areas and enables access to available upstream source headroom. Its benefit is strongly location dependent, with little additional grid side capacity observed at the other evaluated connection points. Data center flexibility provides additional headroom after local network constraints are relieved, with its value depending on workload profile, deferral capability, UPS characteristics, and connection location. The coordinated benefit remains evident under variations in feeder loading, PUE, storage energy, reserve duration, and a combined adverse operating condition. Nonlinear AC power flow replay further supports the physical consistency of the accepted MISOCP solutions. Overall, the results demonstrate that coordinated use of existing grid and data center flexibility can materially increase scenario based feasible interconnection capacity under the evaluated operating conditions, while the achievable benefit remains strongly dependent on local network characteristics and operational assumptions.